\documentclass[onecolumn,sort&compress,numbers]{els-mrw} 

\usepackage{amsmath,amssymb,amsfonts,amsthm,makeidx,graphicx}
\usepackage{txfonts}
\usepackage{helvet}

\begin{document}


\chapter{\boldmath Flavour universality of the $W^\pm$ and $Z$ fermionic couplings}
\label{chap:universality}

\author[1]{Antonio Pich}%


\address[1]{\orgname{IFIC, Universitat de València – CSIC}, 
\orgaddress{Apt. Correus 22085, 46071 València, Spain}}


\maketitle

\begin{abstract}[Abstract]
The Standard Model does not provide any dynamical explanation of the existence of different families of fermions. To account for this experimental fact, it just replicates three times its single-family gauge structure. The equal-charge fermions of the different families couple to the gauge bosons with exactly the same coupling strength. We overview the empirical evidence supporting this important property. The currently most precise experimental tests on the universality of the lepton and quark couplings are discussed. Both charged-current ($W^\pm$) and neutral-current ($Z$) interactions are reviewed.
\end{abstract}

\begin{keywords}
 	Lepton universality\sep quark flavour \sep charged currents\sep neutral currents \sep electroweak interactions 
\end{keywords}


\begin{glossary}[Nomenclature]
	\begin{tabular}{@{}lp{34pc}@{}}
	CKM & Cabibbo-Kobayashi-Maskawa\\
	CL  & Confidence Level\\
	LEP & Large Electron--Positron Collider\\
	LHC & Large Hadron Collider\\
	PDG & Particle Data Group\\
	PMNS & Pontecorvo–Maki–Nakagawa–Sakata\\
	SLC  & SLAC Linear Collider\\
	SM & Standard Model
	\end{tabular}
\end{glossary}

\section*{Objectives}
\begin{itemize}
	\item Overview the family--universal character of gauge couplings within the Standard Model.
	\item Discuss the most precise experimental tests of the universality of the lepton and quark couplings to the $W^\pm$ and $Z$ bosons.
\end{itemize}

\section{Introduction}\label{intro}

The known leptons and quarks constitute the fermionic matter content of the Standard Model (SM), which is organised in a three-fold family structure:
\begin{equation}\label{eq:families}
\left[\begin{array}{cc} \nu_e & u\\ e^- & d'\end{array}\right] ,\qquad
\left[\begin{array}{cc} \nu_\mu & c\\ \mu^- & s'\end{array}\right] ,\qquad
\left[\begin{array}{cc} \nu_\tau & t\\ \tau^- & b'\end{array}\right] .
\end{equation}
The $\mathrm{SU}(2)_L\otimes \mathrm{U}(1)_Y$ gauge theory is defined in terms of the left-handed doublets and right-handed singlets of a single fermion family, which provides an anomaly-free theoretical framework when combined with the $\mathrm{SU}(3)_C$ colour interactions of the corresponding quark components. To account for the existence of several families, the SM just replicates three times the resulting structure. The three fermionic families have identical interactions with the gauge bosons. Fermions with the same electromagnetic charge only differ by their flavour quantum numbers and their masses, i.e., by their couplings to the Higgs boson.

Table~\ref{tab:MassesLifetimes} shows that the measured masses and lifetimes of the three charged leptons span a very broad range of values. The observed mass spectrum manifests a hierarchy of the original Yukawa couplings, which increase from one family to the other. A similar pattern emerges in the quark sector. However, the huge differences of the charged-lepton lifetimes are easily understood as a kinematic reflection of their different masses.

\begin{table}[tbh]
	\TBL{\caption{Measured masses and lifetimes of the three charged leptons \cite{ParticleDataGroup:2024cfk}}}{%
		\begin{tabular*}{\columnwidth}{@{\extracolsep\fill}lll@{}}
			\toprule
			Lepton & Mass & Lifetime\\
			\colrule
			$e$ & $(0.51099895000\pm 0.00000000015)\;\mathrm{MeV}$ &  $> 6.6 \cdot 10^{28}\; \mathrm{yr}\quad$ (90\% CL) \\
			$\mu$ & $(105.6583755\pm 0.0000023)\;\mathrm{MeV}$ & $(2.1969811\pm 0.0000022)\cdot 10^{-6}\; \mathrm{s}$\\
			$\tau$ & $(1776.93\pm 0.09)\;\mathrm{MeV}$ & $(290.3\pm 0.5)\cdot 10^{-15}\; \mathrm{s}$ \\
			\botrule
	\end{tabular*}}{}
	\label{tab:MassesLifetimes}
\end{table}

The tiny neutrino masses ($m_{\nu_i} < 0.8$~eV, 90\% CL) are many orders of magnitude lighter than the very heavy top mass ($m_t\sim 173$~GeV). This huge separation of physical scales is currently not understood. The SM (minimally extended to account for neutrino masses) only relates masses with Higgs interactions, postponing their explanation to the next dynamical level: why are the Yukawa couplings of the different fermions so different? 
The fermionic couplings of the SM Higgs doublet not only distinguish the fermion families, they actually mix them; i.e., the Yukawa couplings are not diagonal in flavour space, which implies that the weak eigenstates $d'_i = d', s', b'$ in Eq.~(\ref{eq:families}) are quantum superpositions of the fermion mass eigenstates $d_j = d, s, b$:
\begin{equation}\label{eq:CKM}
d'_i = \sum_j V_{ij}\, d_j\, ,
\end{equation}
with $V_{ij}$ a $3\times 3$ unitary matrix, known as the Cabibbo-Kobayashi-Maskawa (CKM) matrix. The measured mixings among equal-charge fermion fields
exhibit a very different and unexplained structure in the quark (CKM matrix) and lepton (PMNS matrix) sectors. The fundamental flavour dynamics underlying the numerical values and structure of the fermion masses and mixings is at present totally unknown, reflecting our ignorance about the Higgs interactions.

While we do not really understand the differences among families, we can nevertheless check whether the interactions between the fermions and the electroweak gauge bosons have indeed the same strength for all fermion families. Any observed deviation could provide hints about the unknown dynamics responsible for the flavour structure. 
In the following sections, we discuss the empirical evidence supporting the universality of the electroweak fermion couplings. We will focus mostly on the lepton sector, where the experimental and theoretical accuracies allow for more significant tests, but the available information on the quark sector will also be presented.

\section{Charged-current universality}
\label{sec:CCuniv}

The fermionic couplings of the $W^\pm$ bosons are determined by the charged-current Lagrangian
\begin{equation}\label{eq:Lcc}
\mathcal{L}_{\mathrm{CC}} = -\frac{g}{2\sqrt{2}}\,\left\{ W^\dagger_\mu \,\left[\sum_{ij}\bar u_i\gamma^\mu(1-\gamma_5) V_{ij}d_j +\sum_\ell \bar\nu_\ell\gamma^\mu (1-\gamma_5)\ell\,\right] + \mathrm{h.c.}\,\right\} , 
\end{equation}
and are characterized by a universal coupling $g$. In the original basis of fermion weak eigenstates, quarks and leptons have identical interactions. The unitary mixing matrix $V_{ij}$, which couples up-type and down-type quarks of different families, arises from the redefinition of quark fields in Eq.~(\ref{eq:CKM}), associated with the diagonalization of the quark mass matrices. If neutrino masses are neglected, the analogous mixing matrix in the lepton sector can be eliminated by a redefinition of the neutrino fields.

\begin{figure}[tbh]\centering
\includegraphics[width=10cm]{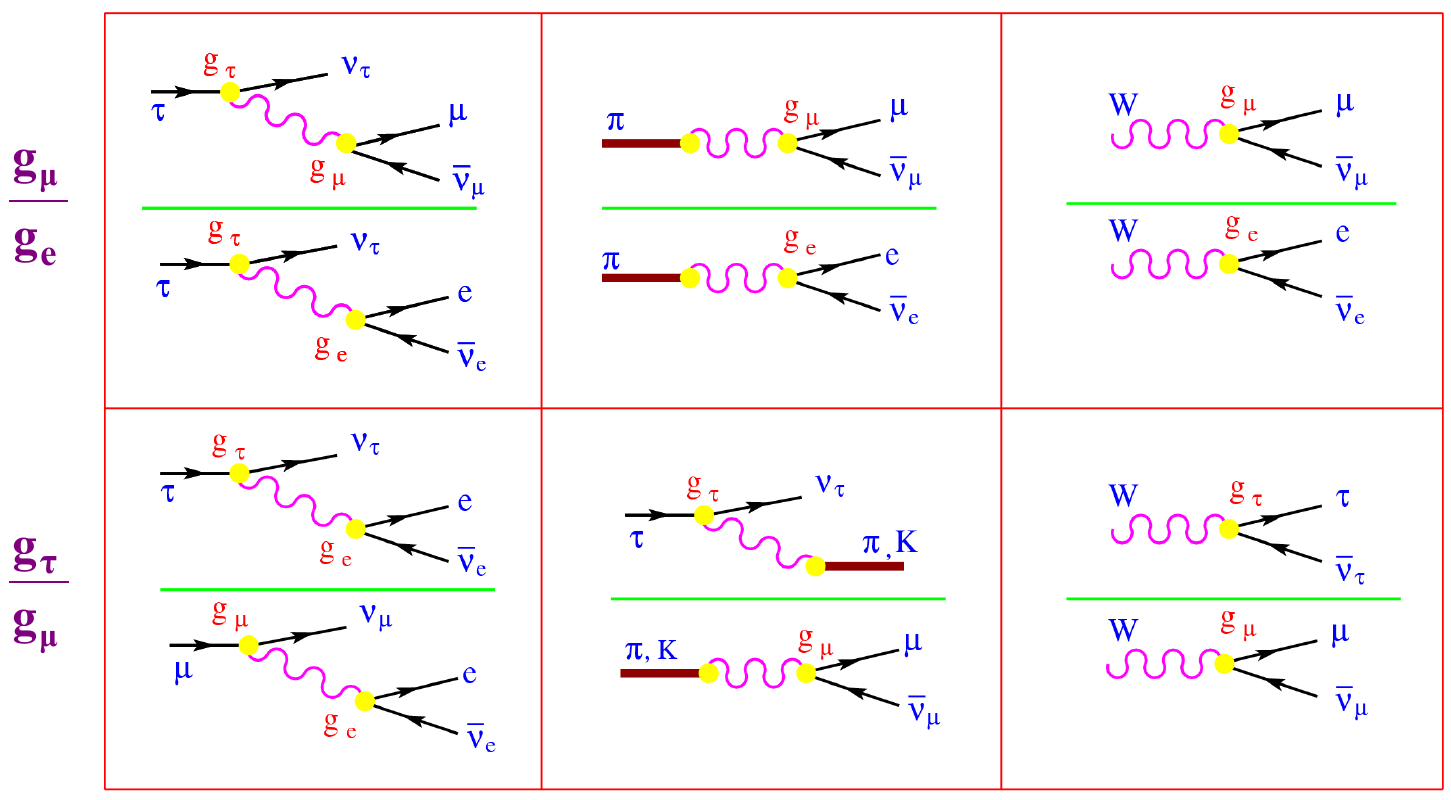}
\caption{Ratios of transition amplitudes sensitive to the ratios of leptonic couplings $g_\ell/g_{\ell'}$.}
\label{fig:LeptonUniversality}
\end{figure}

Let us denote $g_\ell$ the W coupling to the leptonic $\bar\nu_\ell\gamma^\mu (1-\gamma_5)\ell$ current, with $\ell=e,\mu,\tau$. The flavour universality of the leptonic $W^\pm$ couplings, i.e. that $g_e = g_\mu = g_\tau \equiv g$, can be tested by comparing the amplitudes of charged-current transitions that only differ by its lepton content. Figure~\ref{fig:LeptonUniversality} displays a few selected examples, which provide the currently most accurate tests of universality \cite{Pich:2013lsa}. The corresponding determinations of the ratios $g_\ell/g_{\ell'}$ are given in Table~\ref{tab:universality}.

\begin{table}[tbh]
	\TBL{\caption{Most precise experimental determinations of the ratios \ $|g_\ell/g_{\ell'}|$.}}{%
		\begin{tabular*}{\columnwidth}{@{\extracolsep\fill}llllll@{}}
		\toprule  
&
 $\Gamma_{W\to\mu} /\Gamma_{W\to e}$ &
 $\Gamma_{\tau\to\mu}/\Gamma_{\tau\to e}$ &
 $\Gamma_{\pi\to\mu} /\Gamma_{\pi\to e}$ &
 $\Gamma_{K\to\mu} /\Gamma_{K\to e}$ &
 $\Gamma_{K\to\pi\mu} /\Gamma_{K\to\pi e}$  
\\ 
 $|g_\mu/g_e|$
 & $1.000\; (2)$
 & $1.0002\; (11)$ & $1.0010\; (9)$ & $0.9978\; (18)$ & $1.0009\; (18)$ 
\\ \colrule
&
 $\Gamma_{W\to\tau}/\Gamma_{W\to\mu}$ &
 $\Gamma_{\tau\to e}/\Gamma_{\mu\to e}$ &
 $\Gamma_{\tau\to\pi}/\Gamma_{\pi\to\mu}$ &
 $\Gamma_{\tau\to K}/\Gamma_{K\to\mu}$ &
\\ 
 $|g_\tau/g_\mu|$
 & $1.001\; (10)$
 & $1.0016\; (14)$ 
 & $0.996\; (4)$ & $0.986\; (8)$ 
\\ \colrule
 & $\Gamma_{W\to\tau}/\Gamma_{W\to e}$
 &  $\Gamma_{\tau\to\mu}/\Gamma_{\mu\to e}$
\\
 $|g_\tau/g_e|$
 & $1.007\; (10)$
 & $1.0018\; (14)$ 
\\ \botrule
\end{tabular*}}{}\label{tab:universality}
\end{table}

\subsection[$W^-\to\ell^-\bar\nu_\ell$]{\boldmath $W^-\to\ell^-\bar\nu_\ell$}

The partial decay width of the $W^-$ boson into a charged lepton and its corresponding antineutrino,
\begin{equation}
\Gamma_{W\to\ell}\equiv
\Gamma(W^-\to\ell^-\bar\nu_{\ell}) = \frac{g_\ell^2}{48\pi}\,  M_W\;
\left(1 -\frac{3 m_\ell^2}{2 M_W^2} + \frac{m_\ell^6}{2 M_W^6}  \right)\;
\left(1 +\delta^\ell_{\mathrm{RC}}\right)\, ,
\end{equation}
is proportional to $g_\ell^2$. The three leptonic decay modes share the same radiative correction $\delta^\ell_{\mathrm{RC}}$, to a very good approximation. Therefore, the ratio of two leptonic decay widths directly gives the ratio of the squares of the corresponding couplings, up to a very tiny mass correction. From the experimental ratios \cite{ParticleDataGroup:2024cfk}
\begin{equation}
\frac{\Gamma(W^-\to\mu^-\bar\nu_\mu)}{\Gamma(W^-\to e^-\bar\nu_e)} = 1.000\pm 0.004\, ,
\qquad\qquad
\frac{\Gamma(W^-\to\tau^-\bar\nu_\tau)}{\Gamma(W^-\to e^-\bar\nu_e)} = 1.015\pm 0.020\, ,
\qquad\qquad
\frac{\Gamma(W^-\to\tau^-\bar\nu_\tau)}{\Gamma(W^-\to \mu^-\bar\nu_\mu)} = 1.002\pm 0.020\, ,
\end{equation}
one derives the results given in Table~\ref{tab:universality}, which confirm the expected universality of the leptonic charged-current couplings at the 1\% level (0.2\% in the $\mu/e$ case).

This direct test has been significantly improved in recent years, thanks to precise measurements of the ratios of leptonic $W$ branching fractions at LHC \cite{LHCb:2016zpq,ATLAS:2016nqi,ATLAS:2020xea,CMS:2022mhs,ATLAS:2024tlf}. Previous measurements at LEP \cite{ALEPH:2013dgf}, which suffered from much poorer statistics, exhibited a 2.7 $\sigma$ excess in $W^-\to\tau^-\bar\nu_\tau$ decays that was difficult to understand theoretically, given the much stronger constraints from $W$-exchange processes displayed in the table.

\subsection[$\ell^-\to \ell'^-\bar\nu_{\ell'}\nu_\ell$]{\boldmath 
$\ell^-\to \ell'^-\bar\nu_{\ell'}\nu_\ell$}

The leptonic decay widths  of the $\mu$ and the $\tau$ are given by
\begin{equation}
\label{eq:leptonic}
\Gamma_{\ell\to \ell'} \, \equiv \,
\Gamma [\ell^-\to \ell'^-\bar\nu_{\ell'}\nu_\ell (\gamma)]  \, = \,
\frac{g_\ell^2 g_{\ell`}^2}{6144\pi^3}\;\frac{ m_\ell^5}{M_W^4}\;
  f\!\left({m_{\ell'}^2 / m_{\ell}^2}\right) \; \left( 1 + \delta_{\mathrm{\scriptstyle RC}}^{\ell'\ell}\right) \, ,
\end{equation}
where
$\, f(x) = 1-8x+8x^3-x^4-12x^2\log{x}\,$. 
The tiny neutrino masses have been neglected, and $(\gamma)$ indicates additional photons or lepton pairs, which are included inclusively in the radiative correction 
$\delta_{\mathrm{\scriptstyle RC}}^{\ell'\ell}$. For dimensional reasons, the decay width is proportional to the fifth power of the parent lepton mass, which explains the huge difference of seven orders of magnitude between the muon and tau lifetimes.

The lepton-mass dependence of the radiative correction factor $\delta_{\mathrm{\scriptstyle RC}}^{\ell'\ell}$ is very small and can be safely neglected with the current accuracies. Taking the small $f(m_\ell^2/m_\tau^2)$ ($\ell=e,\mu$) correction into account, the experimental ratio \cite{Banerjee:2024znd}
\begin{equation}
\frac{\Gamma_{\tau\to \mu}}{\Gamma_{\tau\to e}} = 0.9730\pm 0.0022\, ,
\end{equation}
directly implies the corresponding $|g_\mu/g_e|$ determination given in the table. The other two ratios of leptonic couplings are obtained from the measured branching ratios
\cite{Banerjee:2024znd}
\begin{equation}
\mathrm{Br}(\tau^-\to\mu^-\bar\nu_\mu\nu_\tau) = 0.1736\pm 0.0004\, ,
\qquad\qquad
\mathrm{Br}(\tau^-\to e^-\bar\nu_e\nu_\tau) = 0.1784\pm 0.0004\, ,
\end{equation}
together with the $\mu$ and $\tau$ lifetimes.
These determinations confirm lepton universality with an impressive accuracy of $1.4\cdot 10^{-3}$.

\subsection{Semileptonic transitions}
\label{subsec:semileptonic}

\subsubsection[$P^-\to\ell^-\bar\nu_\ell$ and $\ell^-\to P^-\nu_\ell$]{\boldmath $P^-\to\ell^-\bar\nu_\ell$ and $\ell^-\to P^-\nu_\ell$}

The decays $P^-\to\ell^-\bar\nu_\ell$ and $\ell^-\to P^-\nu_\ell$ involve one pseudoscalar particle ($P=\pi,K$), either in the initial or final state. The non-perturbative effect of the strong interaction is contained in the hadronic matrix elements of the corresponding quark currents,
\begin{equation}
\langle 0|\bar u\gamma^\mu\gamma_5 d| \pi^-(p)\rangle = i\sqrt{2} F_\pi p^\mu\, ,
\qquad\qquad\qquad
\langle 0|\bar u\gamma^\mu\gamma_5 s| K^-(p)\rangle = i\sqrt{2} F_K p^\mu\, ,
\end{equation}
which are parametrized through the so-called decay constants $F_P$. Taking appropriate 
ratios of transition amplitudes with the same pseudoscalar particle, the dependence on these hadronic parameters factorizes, together with the $V_{ud}$ or $V_{us}$ CKM factor appearing in the quark vertex. Therefore, these ratios can be accurately predicted:
\begin{eqnarray}
R_{P\to e/\mu} &\!\!\!\equiv &\!\!\!
\frac{\Gamma[P^-\to e^-\bar\nu_e (\gamma)]}{
\Gamma[P^-\to\mu^-\bar\nu_\mu (\gamma)]}\; =\; \left|\frac{g_e}{g_\mu}\right|^2\;
\frac{m_e^2}{m_\mu^2}\; \left( \frac{1 - m_e^2/m_P^2}{1 - m_\mu^2/m_P^2}\right)^2\;
\left( 1 +\delta R_{P\to e/\mu}\right)\, ,
\\
R_{\tau/P} &\!\!\! \equiv &\!\!\!
 {\Gamma(\tau^-\to\nu_\tau P^-) \over
 \Gamma(P^-\to \mu^-\bar\nu_\mu)}\; =\;
\Big\vert {g_\tau\over g_\mu}\Big\vert^2\; {m_\tau^3\over 2 m_P m_\mu^2}\;
{(1-m_P^2/ m_\tau^2)^2\over (1-m_\mu^2/ m_P^2)^2}\;
\left( 1 + \delta R_{\tau/P}\right)\, .
\end{eqnarray}

Comparing the experimental ratios \cite{ParticleDataGroup:2024cfk}
\begin{equation}
R_{\pi\to e/\mu} = (1.2327\pm 0.0023)\cdot 10^{-4}\, ,
\qquad\qquad
R_{K\to e/\mu} = (2.488\pm 0.009)\cdot 10^{-5}\, ,
\end{equation}
with the SM predictions \cite{Cirigliano:2007xi,Cirigliano:2007ga,Bryman:2021teu}
\begin{equation}
R_{\pi\to e/\mu}^{\mathrm{SM}} = (1.23524\pm 0.00015)\cdot 10^{-4}\, ,
\qquad\qquad
R_{K\to e/\mu}^{\mathrm{SM}} = (2.477\pm 0.001)\cdot 10^{-5}\, ,
\end{equation}
one extracts the third and fourth determinations of $|g_\mu/g_e|$ in  Table~\ref{tab:universality}. This level of accuracy requires a good theoretical control of the radiative correction factors $\delta R_{P\to e/\mu}$, which are sensitive to the hadronic structure  \cite{Cirigliano:2007xi,Cirigliano:2007ga}.

The present uncertainties on the measured $\tau\to\pi$ (0.5\%) and $\tau\to K$ (1.4\%) branching ratios \cite{Banerjee:2024znd},
\begin{equation}
\mathrm{Br}(\tau^-\to\pi^-\nu_\tau) = 0.1081\pm 0.0005\, ,
\qquad\qquad
\mathrm{Br}(\tau^-\to K^- \nu_\tau) = 0.00696\pm 0.00010\, ,
\end{equation}
dominate the experimental error on the ratios $R_{\tau/\pi}$ and $R_{\tau/K}$. The theoretical accuracy is limited by our current ability to control the hadronic-structure dependence of the radiative corrections.
A recent estimate of these corrections finds 
$\delta R_{\tau/\pi} =(0.18\pm 0.57)\%$ and $\delta R_{\tau/K} =(0.97\pm 0.58)\%$
\cite{Arroyo-Urena:2021nil}, which gives the corresponding $|g_\tau/g_\mu|$ determinations in the table. The experimental uncertainty dominates the error of the kaon determination, while theoretical and experimental uncertainties are of similar size in the pion case.

\subsubsection[$K\to\pi\ell^-\bar\nu_\ell$]{\boldmath $K\to\pi\ell^-\bar\nu_\ell$}

The $K_{\ell 3}$ decay amplitude involves the hadronic matrix element of the vector quark current between the initial kaon and the final pion:
\begin{equation}
\langle \pi(p_\pi)|\,\bar u\gamma^\mu s\, | K(p_K)\rangle = C_{K\pi}\,\left\{
\left[ (p_K+p_\pi)^\mu - \frac{m_K^2-m_\pi^2}{t}\, (p_K-p_\pi)^\mu\right]\, f^{K\pi}_+(t)
+ \frac{m_K^2-m_\pi^2}{t}\, (p_K-p_\pi)^\mu\, f^{K\pi}_0(t)\right\}\, ,
\end{equation}
where $C_{K^-\pi^0} = 1/\sqrt{2}$, $C_{\bar K^0\pi^+}=1$,
$t = (p_K-p_\pi)^2=(p_\ell+ p_{\bar\nu_\ell})^2$ is the momentum transfer to the leptonic $\ell^-\bar\nu_\ell$ pair and $ f^{K\pi}_+(0)= f^{K\pi}_0(0)$. The contribution from the scalar form factor $f^{K\pi}_0(t)$ to the decay rate is suppressed by a factor $m_\ell^2$. Taking out a global factor $|V_{us}|^2 |f^{K\pi}_+(0)|^2$, from the measured differential distribution it is possible to extract the ratios $|f^{K\pi}_+(t)/f^{K\pi}_+(0)|$ and $|f^{K\pi}_0(t)/f^{K\pi}_+(0)|$, and compute their contributions to the phase-space integral. Combining this information with the total decay width, one can then determine experimentally the global factor $|V_{us} f^{K\pi}_+(0)|^2$ \cite{Antonelli:2009ws} (times the square of the product of charged-current couplings at the leptonic and quark vertices). 
This global factor and short-distance radiative corrections cancel out in the ratio between the results obtained from the muonic and electronic rates.
Taking into account the computed long-distance radiative corrections \cite{Cirigliano:2008wn,Seng:2021boy}, one gets then a determination of $|g_\mu/g_e|$.
From $K_L$ decays, this ratio is found to be $1.0022\pm 0.0024$, while $K^\pm$ decays give $0.9995\pm 0.0026$ \cite{Bryman:2021teu,Seng:2021nar}. The average of these two determinations leads to the final entry in the first row of Table~\ref{tab:universality}. 

\subsubsection{Constraints from heavy-quark decays}

Lepton universality can also be tested in decays of charm and bottom, although at a much less precise level. Owing to their strong helicity suppression by the $m_\ell^2/m_{D,B}^2 $ mass ratio, the decays $D^-\to e^-\bar\nu_e$, $B^-\to e^-\bar\nu_e$  and $B^-\to \mu^-\bar\nu_\mu$ have not been measured yet, while the ratio $R_{D\to\tau/\mu}$ is only known experimentally with a 23\% error. On the other hand, the current theoretical uncertainties of the relevant hadronic form factors limit the possible accuracy of $|g_\ell/g_{\ell'}|$ determinations from semileptonic decays with additional hadrons in the final state.

In 2012 the BaBar collaboration reported an excess of $\bar B\to D^{(*)}\tau^-\bar\nu_\tau$ events \cite{BaBar:2012obs}: the ratios $R(D^{(*)})\equiv \Gamma(\bar B\to D^{(*)}\tau^-\bar\nu_\tau)/\Gamma(\bar B\to D^{(*)}\ell^-\bar\nu_\ell)$ with $\ell=e,\mu$ were found to be $2.0\,\sigma$ and $2.7\,\sigma$ larger than their SM expectations for $R(D)$ and $R(D^*)$, respectively, giving a combined anomaly of $3.4\,\sigma$. Since then, these ratios have been measured by Belle, LHCb and BelleII, which find them in better agreement with the SM (the most recent BelleII measurements agree with the SM predictions within $1.5\,\sigma$). Nevertheless, the weighted average of all measurements still exhibits a combined $3.8\,\sigma$ anomaly \cite{Banerjee:2024znd,HFLAV:2025ckm}. Additional studies are needed to clarify the situation.

\subsection{Physics implications}

The naive average of the determinations in Table~\ref{tab:universality} gives
\begin{equation}
|g_\mu/g_e| = 1.0003\pm 0.0006\, ,\qquad\qquad
|g_\tau/g_\mu| = 1.0006\pm 0.0013\, ,\qquad\qquad
|g_\tau/g_e| = 1.0019\pm 0.0014\, .
\end{equation}
Thus, the leptonic $W^\pm$ couplings are indeed universal within a precision of around one per mil. However, the separate ratios quoted in the table contain much richer information because each individual test is sensitive to different types of physics beyond the SM. For instance, while $\Gamma_{W\to\ell}/\Gamma_{W\to\ell'}$ directly probes the couplings of an on-shell $W^\pm$, the ratios of leptonic decay widths 
are sensitive to the virtual exchange of a transversely polarized $W^\pm$ boson, whereas $R_{P\to e/\mu}$ tests the longitudinal polarization of the off-shell $W^\pm$ propagator.

In low-energy processes mediated by $W^\pm$ exchange such as $\ell^-\to \ell'^-\bar\nu_{\ell'}\nu_\ell$, $\tau^-\to \nu_\tau d\bar u$ ($\tau^-\to \nu_\tau \pi^-$) or $d\bar u\to\ell^-\bar\nu_\ell$ ($\pi^-\to \ell^-\bar\nu_\tau $), we are actually testing a local four-fermion interaction. Since the momentum transfer is much smaller than $M_W$, the $W$ propagator reduces to a constant, and the distance between the two vertices in Fig.~\ref{fig:LeptonUniversality} can no longer be resolved. For instance, the leptonic decay amplitude is given by
\begin{equation}
T_{\ell\to\ell'} = \frac{1}{8}\, g_\ell g_{\ell'}\, [\bar\ell'\gamma^\mu(1-\gamma_5)\nu_{\ell'}]\, \frac{-g_{\mu\nu}+ q_\mu q_\nu/q^2}{q^2-M_W^2}\,
[\bar\nu_\ell\gamma^\nu (1-\gamma_5)\ell]
\quad\overset{q^2\ll m_W^2}{\longrightarrow}\quad
\frac{g_\ell g_{\ell'}}{8M_W^2}\, [\bar\ell'\gamma^\mu(1-\gamma_5)\nu_{\ell'}]\,
[\bar\nu_\ell\gamma_\mu(1-\gamma_5)\ell]\, .
\end{equation}
Keeping higher orders in the expansion in powers of $q^2/M_W^2$ of the $W$ propagator would generate higher-dimensional operators suppressed by additional powers of the $W$ mass that can be safely neglected.
In the presence of new physics, the transition could also be mediated by a hypothetical scalar boson, which would leave its imprint in a different structure of the four-lepton operator.

One can perform a general analysis of these transitions by writing an effective Lagrangian containing all relevant four-fermion operators. Requiring Lorentz invariance, the leading dimension-six Lagrangian contains vector, scalar and tensor interactions with all possible flavour combinations of left-handed and right-handed fermions. Using the ratios in Table~\ref{tab:universality} and additional experimental information, one can then constrain the low-energy couplings of the effective Lagrangian \cite{Pich:2013lsa,Bryman:2021teu,Cirigliano:2021yto}. If a clear deviation from the SM is experimentally established in a particular transition, this type of analysis could uncover the underlying dynamics that generate the anomalous behaviour.

\section{Neutral-current universality}

The $Z$ boson couples to the SM fermions through the neutral-current Lagrangian
\begin{equation}
\mathcal{L}_{\mathrm{NC}} = -\frac{g}{2\cos{\theta_W}}\, Z_\mu\;\sum_f \bar f\gamma^\mu (v_f-a_f\gamma_5) f\, ,
\end{equation}
where $g$ is the same $\mathrm{SU}(2)_L$ coupling appearing in the charged-current Lagrangian (\ref{eq:Lcc}) and $\theta_W$ is the weak mixing angle ($e = g \sin{\theta_W}$). The interaction is diagonal in flavour and all fermions with equal electric charge have identical vector and axial-vector couplings:
\begin{equation}\label{eq:v_a_Zcouplings}
v_f = T_3^f\, \left( 1 - 4 |Q_f|\sin^2{\theta_W}\right)\, ,
\qquad\qquad
a_f = T_3^f\, ,
\end{equation}
with $T_3^f = +\frac{1}{2}$ or $-\frac{1}{2}$ for the upper or lower components, respectively, of the doublets in Eq.~(\ref{eq:families}).

The partial $Z$ decay width into a final fermion-antifermion state is given by
\begin{equation}
\Gamma(Z\to \bar f f)\, =\, N_f\, \frac{g^2 M_Z}{48\pi\cos^2{\theta_W}}\,\left( |v_f|^2+|a_f|^2\right)\, \left(1+\tilde\delta_{\mathrm{RC}}\right) 
\, =\,
N_f\, \frac{G_F M_Z^3}{6\pi\sqrt{2}}\,\left( |v_f|^2+|a_f|^2\right)\, \left(1+\delta_{\mathrm{RC}}^Z\right)\, ,
\end{equation}
where $N_\ell = 1$ for leptons ($\ell=e,\mu,\tau$), while for quarks the effective factor $N_q = N_C\, (1 + \delta_{\mathrm{QCD}})$ takes into account the number of colours $N_C=3$  and the QCD correction $\delta_{\mathrm{QCD}}$. In the right-hand-side expression the result has been written in terms of the $Z$ mass and the Fermi coupling measured in $\mu$ decay, $G_F = \sqrt{2} g^2/(8M_W^2)$, making use of the relation $M_W^2 = M_Z^2\cos^2{\theta_W}$. This reabsorbs some electroweak radiative corrections and has the advantage of using the most precisely measured parameters to fix the interaction.

The vector and axial-vector couplings can be disentangled by measuring the forward-backward asymmetry in $e^+e^-\to \gamma,Z\to\bar f f$. At the $Z$ peak, where $Z$ exchange completely dominates,
\begin{equation}
\mathcal{A}_{\mathrm{FB}}^f\equiv\frac{N_F-N_B}{N_F+N_B} = \frac{3}{4}\,\mathcal{P}_e\mathcal{P}_f \, .
\end{equation}
$N_F$ and $N_B$ denote the number of $f$'s emerging in the forward and backward hemispheres, respectively, with respect to the electron direction, and
\begin{equation}
\mathcal{P}_f = \frac{-2v_f a_f}{v_f^2+a_f^2}
\end{equation}
is the average longitudinal polarization of the fermion $f$, which only depends on the ratio of the vector and axial-vector couplings. Since $\sin^2{\theta_W}\sim 0.23$, small higher-order corrections can produce large variations on the lepton polarization because
$|v_\ell| =\frac{1}{2}\, |1 - 4\sin^2{\theta_W}|\ll 1$. Therefore, $\mathcal{P}_\ell$ is very sensitive to electroweak quantum effects. This final lepton polarization has only been measured for the $\tau$ lepton, taking advantage of the fact that the spin polarization of the $\tau$ is reflected in the distorted distribution of its decay products.

With polarised $e^+e^-$ beams, which were available at the SLAC linear collider (SLC), one can also measure the asymmetry between the cross sections for initial left- and right-handed-polarised initial electrons, and the corresponding forward-backward left-right asymmetry. At the $Z$ peak they take the values
\begin{equation}
\mathcal{A}_{\mathrm{LR}}^f = \frac{\sigma^f_L-\sigma^f_R}{\sigma^f_L+\sigma^f_R} = -\mathcal{P}_e\, ,
\qquad\qquad\qquad
\mathcal{A}_{\mathrm{FB,LR}}^f = -\frac{3}{4}\,\mathcal{P}_f \, .
\end{equation}
Thus, $\mathcal{A}_{\mathrm{LR}}^f$ measures the longitudinal polarization of the initial electron, while $\mathcal{A}_{\mathrm{FB,LR}}^f$ determines the polarization of the final fermion.

\begin{figure}[tbh]\centering
\includegraphics[width=7cm]{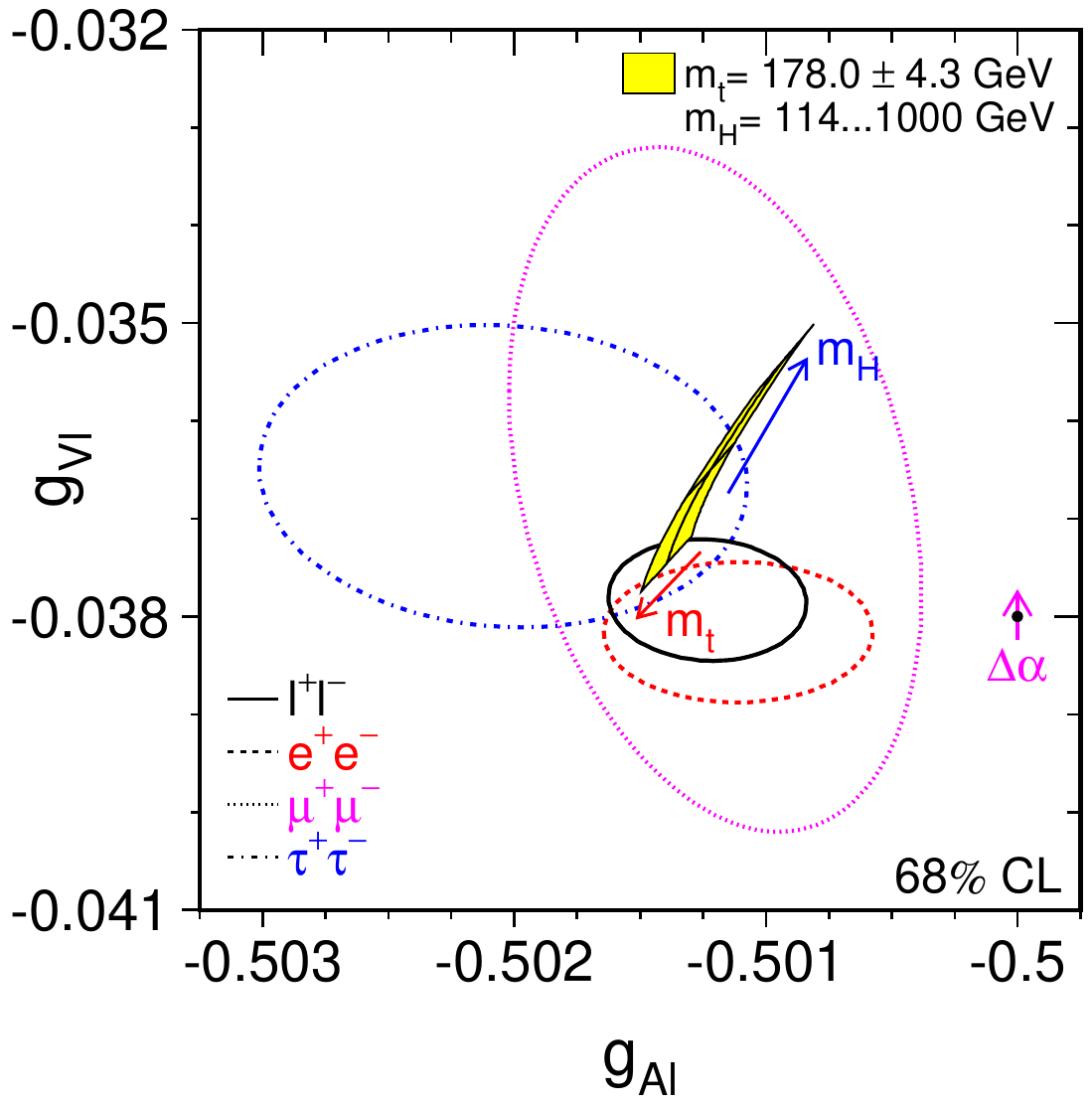}
\caption{Combined LEP and SLC measurements of the effective vector and axial-vector $Z$ couplings \cite{ALEPH:2005ab}.}
\label{fig:NCuniversality}
\end{figure}

The $Z$ couplings were thoroughly investigated in the 1990s at the $Z$ factories LEP-I and SLC. The combined analysis of their full data samples provided the lepton-universality test summarized in Fig.~\ref{fig:NCuniversality} \cite{ALEPH:2005ab}. The coloured ellipses  display the measured axial ($g_{Al} = a_\ell$) and vector ($g_{Vl} = v_\ell$) $Z$ couplings for electrons (red, dashed), muons (violet, dotted) and taus (blue, dash-dotted), at the 68\% confidence level (CL). The three determinations are in good agreement, and their combination provides the more accurate black contour that assumes universality. The crucial role of electroweak radiative corrections is illustrated with the point on the right, labelled with $\Delta \alpha$, which corresponds to the SM prediction with all higher-order corrections removed except the photon vacuum polarization; the arrow displays the error induced by the uncertainty on $\alpha(M_Z^2)$, the QED coupling at the scale $M_Z$. The complete SM prediction is given by the yellow region as a function of the top and Higgs masses, with the arrows pointing in the direction of increasing values of these masses. Although the Higgs boson was not yet discovered, its indirect manifestation through electroweak radiative corrections was already 
indicating that $M_H\sim 120$~GeV should be preferred.

The 68\% CL regions in the figure can be translated into the lepton-universality ratios
\cite{ParticleDataGroup:2024cfk,ALEPH:2005ab}
\begin{equation}\label{eq:NCva}
\frac{v_\mu}{v_e} = 0.961\pm 0.061\, ,
\qquad\qquad
\frac{a_\mu}{a_e} = 1.0002\pm 0.0013\, ,
\qquad\qquad
\frac{v_\tau}{v_e} = 0.959\pm 0.029\, ,
\qquad\qquad
\frac{a_\tau}{a_e} = 1.0019\pm 0.0015\, .
\end{equation}
Thus, universality has been tested at the per-mil level for the axial couplings, while only a few per-cent accuracy has been reached for the smaller vector couplings. These numbers have remained unchanged since the end of the LEP and SLC operations. 

Subtracting from the total $Z$ width the measured partial widths into quarks and charged leptons, one can determine the decay width of the $Z$ boson into invisible modes.
Assuming that this invisible width originates from $N_\nu$ light neutrino species with SM interactions, one determines $N_\nu = 2.9963\pm 0.0074$ \cite{ParticleDataGroup:2024cfk}.
Fixing instead $N_\nu=3$, this can be translated into a measurement of 
\begin{equation}
\frac{|g_{\nu_e}|^2 + |g_{\nu_\mu}|^2 +|g_{\nu_\tau}|^2}{3\, |g_\nu^{\mathrm{SM}}|^2} = 0.9988\pm 0.0025\, .
\end{equation}
A direct flavour-dependent measurement of the $Z$ neutrino couplings from $\nu_\mu e^-$ and $\nu_e e^-$ scattering was performed a long time ago by the CHARM-II experiment \cite{CHARM-II:1993xmq}. From the slightly re-evaluated numbers quoted by the PDG \cite{ParticleDataGroup:2024cfk}, one gets
\begin{equation}
\left|\frac{g_{\nu_\mu}}{g_{\nu_e}}\right| = 0.95\pm 0.16\, ,
\end{equation}
in agreement with universality within its very large uncertainty.

\section{Universality of the quark couplings}

Tests of universality in the quark sector face the difficulties associated with identifying the light quark flavours at high energies. LEP and SLC provided accurate measurements of
the $Z$ boson partial
decay widths into charm and bottom quarks (the $Z$ decay into $\bar t t$ is kinematically forbidden), normalised to its total hadronic width \cite{ParticleDataGroup:2024cfk},
\begin{equation}\label{eq:RcRb}
R_c \equiv \frac{\Gamma(Z\to \bar c c)}{\Gamma(Z\to\mathrm{hadrons})} =
0.1721\pm 0.0030\, ,
\qquad\qquad\qquad
R_b \equiv \frac{\Gamma(Z\to \bar b b)}{\Gamma(Z\to\mathrm{hadrons})} =
0.21629\pm 0.00066\, ,
\end{equation}
in good agreement with the SM predictions. Assuming the SM $Z$ couplings in Eq.~(\ref{eq:v_a_Zcouplings}), this information is used in the global electroweak fit that determines $\sin^2{\theta_W}$. The measurements of decays into light flavours are not so precise. The PDG only quotes averages for the $Z$ decay widths into `up-type' or `down-type'
quarks \cite{ParticleDataGroup:2024cfk}:
\begin{equation}\label{eq:RucRdsb}
R_{uc}\equiv \frac{\Gamma[Z\to (\bar u u+\bar c c)/2]}{\Gamma(Z\to\mathrm{hadrons})} =
0.166\pm 0.009\, ,
\qquad\qquad\qquad
R_{dsb}\equiv \frac{\Gamma[Z\to (\bar d d+\bar s s +\bar b b)/3]}{\Gamma(Z\to\mathrm{hadrons})} =
0.223\pm 0.006\, .
\end{equation}
From the experimental ratios in Eqs.~(\ref{eq:RcRb}) and (\ref{eq:RucRdsb}), one gets (neglecting correlations):
\begin{equation}
\left(\frac{|v_u|^2+|a_u|^2}{|v_c|^2+|a_c|^2}\right)^{1/2} = 0.96\pm 0.06\, ,
\qquad\qquad\qquad
\left(\frac{(|v_d|^2+|a_d|^2)+(|v_s|^2+|a_s|^2)}{2\, (|v_b|^2+|a_b|^2)}\right)^{1/2} = 1.02\pm 0.02\, .
\end{equation}
These results confirm the universality of the neutral-current quark couplings at the few per-cent level.

The $W^\pm$ couplings to the $\bar u_i\gamma^\mu (1-\gamma_5) d_j$ current involve the product $g V_{ij}$ with $g$ the $\mathrm{SU}(2)_L$ gauge coupling and $V_{ij}$ the corresponding element of the unitary CKM matrix. Instead of taking family-dependent couplings $g_i$ ($i=u,c,t$), we can adopt the normalization $g=g_\mu$ and reabsorb any possible violations of universality in $V_{ij}$; i.e., $V_{ij} = g_i V^0_{ij}/g_\mu$ with $V^0_{ij}$ the original unitary matrix. Obviously, if any of the $g_i$ couplings  is not identical to $g_\mu$, this would manifest as a violation of unitarity in the effective $V_{ij}$ matrix.
We can safely assume the universality of the leptonic couplings because, as shown in section~\ref{sec:CCuniv}, they have already been tested to a much higher precision.

The measured leptonic branching fraction of the $W^\pm$ boson determines the sum of the six $|V_{ij}|^2$ factors corresponding to its open quark decay modes \cite{ParticleDataGroup:2024cfk}:
\begin{equation}
\sum_{j=d,s,b} \left(|V_{uj}|^2 + |V_{cj}|^2\right) = 2.002\pm 0.027\, .
\end{equation}
This result is in perfect agreement with the unitarity expectation for the sum of squared entries in the first two rows of the CKM matrix ($\sum_j |V_{ij}|^2 = 1$).

The individual entries of the CKM matrix (except $V_{td}$ and $V_{ts}$) are extracted from semileptonic weak transitions of the type discussed in section~\ref{subsec:semileptonic}, fixing $G_F$ from $\mu$ decay and using lattice calculations of the relevant hadronic parameters such as $F_P$ of $f_+^{K\pi}(0)$. These tree-level determinations give us three additional universality tests, corresponding to the unitarity relations associated with the first two rows and the last column of the quark mixing matrix. Using the values quoted in the most recent PDG compilation \cite{ParticleDataGroup:2024cfk}, one gets:\footnote{We have taken into account a recent next-to-leading-logarithmic QCD analysis of mixed electromagnetic corrections \cite{Moretti:2025qxt}, which slightly increases the value of $|V_{ud}|$ extracted from superallowed nuclear $\beta$ decays from $0.97367 \pm 0.00032$ to $0.97384 \pm 0.00029$.}
\begin{equation}
\sum_j |V_{uj}|^2 = 0.9987\pm 0.0007\, , 
\qquad\qquad
\sum_j |V_{cj}|^2 = 1.001\pm 0.012\, ,
\qquad\qquad
\sum_i |V_{ib}|^2 = 1.02\pm 0.05\, .
\end{equation}
There is a $1.9\,\sigma$   
tension with unitarity in the first row, due to a recent reduction of the value of $V_{ud}$ from superallowed nuclear $\beta$ decays, that needs further clarification. 
All other tests are consistent with unitarity and, therefore, with the expected universality of the charged-current quark couplings.

\section{Conclusions}
\label{sec:conclusions}

The flavour structure is one of the main pending questions in our understanding of the fundamental interactions among the constituents of matter. We have learned experimentally the existence of three families of fermions. Moreover, additional SM families are excluded by the measured Higgs production cross section at the LHC (a fourth family with heavier quarks not yet detected would increase the cross section by a factor of nine). However, we completely ignore the reasons for this replication. In the SM, the three fermionic families have identical gauge interactions. All differences among families emerge from their Yukawa couplings to the Higgs doublet that generate the observed masses and mixings, but the SM does not determine the numerical values of these Yukawa interactions. 

So far, the available data confirm the universal character of the leptonic couplings to the $W^\pm$ and $Z$ bosons. The charged-current interactions of the $e$, $\mu$ and $\tau$ have been verified to be identical with a precision of around one per mil. A similar precision has been reached for the axial couplings to the $Z$ boson, while the smaller vector couplings have only been tested at the few per-cent level. The family independence of the quark couplings is more difficult to test, but, nevertheless, we have empirical evidence that the neutral couplings of the bottom quark agree with the average of the down and strange ones at the 2\% level, while the charm and up ones are equal within a 6\% accuracy. The universality of the charged-current quark couplings is also verified at the few per-cent level by the usual tests of CKM unitarity. 

There is large room for improvements. Ongoing and future experiments will scrutinize these fermionic couplings to a much deeper level of sensitivity, trying to uncover possible signals of the unknown dynamics responsible for the observed flavour structure.

\begin{ack}[Acknowledgments]%
This work is supported by the Spanish Government and ERDF/EU (Agencia Estatal de Investigación MCIU/AEI/10.13039/501100011033),
Grants No. PID2023-146220NB-I00 and CEX2023-001292-S.
\end{ack}


\bibliographystyle{Numbered-Style} 
\bibliography{LUreferences}

\end{document}